\documentclass[hyper]{JHEP} 

\def\be{\begin{equation}}
\def\ee{\end{equation}}
\def\ba{\begin{array}}
\def\ea{\end{array}}
\def\bea{\begin{eqnarray}}
\def\eea{\end{eqnarray}}
\def\nn{\nonumber\\}

\def\ct{\cite}
\def\la{\label}
\def\eq#1{Eq. (\ref{#1})}

\def\a{\alpha}
\def\b{\beta}

\def\D{\Delta}

\def\ph{\phi}

\def\ps{\psi}

\def\k{\kappa}
\def\l{\lambda}

\def\n{\nu}
\def\th{\theta}

\def\r{\rho}
\def\s{\sigma}

\def\ta{\tau}

\def\pr{\prime}

\def\half{\frac{1}{2}}

\def\pa{\partial}

\def\lb{\left[}
\def\lc{\left\{}
\def\ls{\left(}
\def\lp{\left.}
\def\rp{\right.}
\def\rb{\right]}
\def\rc{\right\}}
\def\rs{\right)}

\newcommand\fverb{\setbox\pippobox=\hbox\bgroup\verb}

\newcommand\fverbdo{\egroup\medskip\noindent%

            \fbox{\unhbox\pippobox}\ }

\newcommand\fverbit{\egroup\item[\fbox{\unhbox\pippobox}]}

\newbox\pippobox

\title{Spiky Strings in AdS$_4 \times$ {\bf CP}$^3$ with
Neveu-Schwarz Flux}

\author{Sachin Jain $^a$ and Kamal L. Panigrahi $^b$\\
\vspace{1cm}
$^a$ Institute Of Physics, Bhubaneswar, 751 005, India \\
$^b$ Department of Physics, Indian Institute of Technology Guwahati,\\
~~Guwahati-781 039, India

\vspace{1cm}

E-mail: \email{sachjain@iopb.res.in, panigrahi@iitg.ernet.in}}

\preprint{}

 \abstract{We study general rotating string solution in the
$AdS_4\times {\bf CP}^3$ background with a $B_{NS}$ holonomy turned on
over ${\bf CP}^1$ $\subset $ $ {\bf CP}^3$. We find the giant magnon and single spike solutions
for the string moving in this background corresponding to open spin chain.
We calculate the corresponding dispersion relation among various conserved
charges for both the cases. We further study the finite size effect
on both the giant magnon and single spike solutions.}


\begin{document}

\section{Introduction}\label{first}
Recently Aharony, Bergman, Jafferis and Maldacena (ABJM)
\cite{ABJM} proposed a new class of gauge theory-string theory duality
between ${\cal N} = 6$ Chern-Simons (CS) theory and type IIA string
theory on AdS$_4 \times {\bf CP}^3$. Based on this Aharony,
Bergman, and Jafferis (ABJ) \cite{Aharony:2008gk} identified further a class
of $AdS_4/CFT_3$ duality with extended supersymmetry namely, a
three dimensional ${\cal N}$ = 6 superconformal Chen-Simons
theory with a gauge group $U(M)_k \times {\overline{U(N)}}_{-k}$,
with $k$ being the level of the CS theory, is dual to type IIA
string theory on AdS$_4 \times {\bf CP}^3$ with B$_{NS}$ holonomy
turned on over {\bf CP}$^1$ $\subset$ {\bf CP}$^3$.

In proving AdS/CFT duality \cite{MGW}, the integrability of both the string
theory and the gauge theories have played a key role. The
semiclassical string states in the gravity side has been used to
look for suitable gauge theory operators on the boundary, in
establishing the duality. In this connection, Hofman and Maldacena
(HM) \ct{HM} considered a special limit where the
problem of determining the spectrum of both sides becomes rather
simple. The spectrum consists of an elementary excitation known as
magnon which propagates with a conserved momentum $p$ along the
spin chain. Further, a general class of rotating string
solution in AdS$_5$ is the spiky string
\cite{Kruczenski:2004wg,krt0607} which describes the higher twist
operators from dual field theory view point. Giant magnons can be
thought of as a special limit of such spiky strings with shorter
wavelength. Recently there has been a lot of work devoted
for the understanding of
the giant magnon and spiky string solutions in various backgrounds
see for example
\ct{BIKS}-\ct{KNP}.
There has also been
numerous papers devoted for understanding the finite size corrections
on these solutions
\ct{AFZ}.

In the present paper we study a general class of rotating string
solution in the ABJ model. It has already been established that
like its regular counter part, ABJ theory has an integrability
structure in planar limit. In fact in \cite{BGR} it has
further been analyzed that the magnon dispersion relation remains
exactly the same as that of ABJM, even the ABJ theory has an extra NS B
field in its spectrum, thus showing no parity symmetry breaking
effect. In this paper we study the giant magnon and spike solutions
for strings
corresponding to open spin chain. We concentrate on a
particular sector of the theory which is the diagonal $SU(2)$
subgroup inside ${\bf CP}^3$, and study a general class
of rotating strings. We solve the equations of motion and the
Virasoro constraints for the Polyakov action of the string in the
presence of a NS-NS $B$ field. We find out the general form of all
the conserved charges and choose particular parametrization
corresponding to a special relation among those charges. We further
find the corresponding dispersion relation for the giant magnon
and single spike solutions for the string moving in the $SU(2)$
subsector inside ${\bf CP}^3$. Finally, we study the finite size
corrections for both the giant magnon and for the single spike
solutions.

The rest of the paper is organized as follows. In section 2, we
consider a rotating string solution in $R \times S^2$
with NS-NS $B$ field. We
write down all the equations of motion, and
Virasoro constraints for the string moving in this background. In
section 3, we find the giant magnon and single spike solutions for the
string in infinite size limit and the corresponding dispersion
relations inside a particular parameter space. In
section 4, we investigate the finite size effects for both the
giant  magnon and single spike solution.
In section 5, we present our conclusions.

\section{Rotating String Solutions in ABJ theory}
We start by writing down the supergravity dual background of the
ABJ theory \bea ds^2 &=& \frac{1}{4} R^2 \lb - \cosh^2 \r \ dt^2 +
d\r^2 + \sinh^2 \r \ls d \eta^2 + \sin^2 \eta d \chi^2 \rs \rb \nn &&
\quad + R^2 \lb d\xi^2 + \cos^2 \xi \sin^2 \xi \ls d \ps + \half
\cos \th_1 d \ph_1 - \half \cos \th_2 d \ph_2 \rs^2 \rp \nn && \ \
\lp + \frac{1}{4} \cos^2 \xi \ls d \th_1^2 + \sin^2 \th_1 d
\ph_1^2 \rs + \frac{1}{4} \sin^2 \xi \ls d \th_2^2 + \sin^2 \th_2
d \ph_2^2 \rs \rb . \nn B_{\rm NS}&=& -{B\over 2} \Bigl( {\sin
2\xi } {\rm d} \xi \wedge \left(
 2{\rm d }\psi + {\cos\theta_1} {\rm d} \phi_1 -
{\cos\theta_2 } {\rm d} \phi_2 \right)\Bigr. \nn && \ \ \ \
\Bigr.+ \cos^2\xi \sin\theta_1 {\rm d} \theta_1 \wedge {\rm d}
\phi_1 + \sin^2\xi \sin\theta_2 {\rm d} \theta_2 \wedge {\rm d}
\phi_2 \Bigl) \, .\eea In addition to this there is a dilaton
field and Ramond-Ramond two form and four form fields, whose
detailed forms will not be needed in what follows. When taking
$\a^{\pr} =1$, the curvature radius $R$ is given by $R^2 = 2^{5/2}
\pi \l^{1/2}$, which is precisely same as that of ABJM. We are
interested in a particular sector this model which can be obtained
by choosing $\r = 0$, $\ps$ = constant and $\xi =\frac{\pi}{4} $
and by identifying
 that $\th_1 = \th_2 \equiv \th$ and $\ph_1 = \ph_2
\equiv \ph$. With the above identifications, AdS$_4 \times$ ${\bf
CP}^3$ geometry reduces to $R \times S^2$ \footnote{From the
general equations of motion for the string moving inside the full
$AdS_4 \times {\bf CP}^3$, one can check that the above
identification also gives a consistent solution}. The metric and
the NS-NS flux reads \bea \la{redmet} ds^2 &=& \frac{1}{4} R^2 \lb
-dt^2  + \ls d \th^2 + \sin^2 \th d \ph^2 \rs\rb , \nn B_{\rm NS}
&=& -{B\over 2} \left( \sin\theta {\rm d} \theta\wedge {\rm d}
\phi\right) . \eea

We are interested in studying the classical rotating string around
this geometry. We use the Polyakov action \bea\label{actPol} S =
\frac{1}{4 \pi} \int d\sigma d\tau
[\sqrt{-\gamma}\gamma^{\alpha\beta} g_{MN}\partial_\alpha
x^M\partial_\beta x^N + e^{\alpha\beta}
\partial_\alpha x^M\partial_\beta x^N B_{MN}] \nn
\eea where we choose $\gamma^{\a\b} = \eta^{\a\b}$ as the world
sheet metric and $e^{\alpha\beta}$ is the anti symmetric tensor
defined as $e^{01}=-e^{10} = 1$. In terms of target space
coordinates the action is given by
\bea S &=& \frac{\pi \sqrt{2\l}}{4\pi} \int d \s d\tau
    \lb
    (\pa_{\ta} t)^2 - (\pa_{\s} t)^2 - (\pa_{\ta} \th)^2 + (\pa_{\s} \th)^2 - \sin^2 \th
        \lc
        (\pa_{\ta} \ph)^2 -(\pa_{\s} \ph)^2
        \rc
    \rb \nn
 &+& \frac{B}{4 \pi}\int d\s d\tau \sin\theta
    \lb
    (\pa_{\s}\th \pa_{\ta} \ph) - (\pa_{\ta}\th \pa_{\s} \ph) .
    \rb
\eea To find the spiky string in this geometry,
 we choose the following parametrization \bea t = f(\ta)
,\>\> \th = \th (y), \>\> \phi = \n \ta + h (y) , \eea where $y =
a \ta + b \s$ and we set $\ta$ and $\s$ run from $-\infty$ to
$\infty$. Looking at the background geometry, one infers that
there exist two conserved charges and the equations of motion for
the corresponding fields are given by, \bea 0 &=& \pa_{\ta}^2
f(\ta) , \nn 0 &=& \pa_y \lb \sin^2 \th \lc a \n +(a^2 - b^2)
h^{\pr} \rc \rb , \eea where prime implies derivative with
respect to $y$. The solutions of these equations are \bea f(\ta)
&=& \k \tau \ , \nn h'(y) &=& \frac{1}{a^2 -b^2} \frac{c - a\n
\sin^2 \th }{\sin^2 \th} \ , \eea where $\k$ and $c$ are
integration constants. Further the Virasoro constraints $T_{\a\b} = 0$
have also to be imposed.
Due to the symmetry property of the metric, and the
conformal nature of the Polyakov action
the number of real independent constraints are two and we write them
in the following form for our convenience
\bea T_{\ta\ta} + 2 T_{\ta\s} = 0 \ , \>\>\>
T_{\ta\ta} - \frac{a^2 + b^2}{2 a b} T_{\ta\s} = 0. \eea The first
Virasoro constraint gives rise to  \be \la{the} \th' = \frac{b
\n}{a^2 -b^2} \frac{\sqrt{ (\sin^2 \th_{max} - \sin^2 \th) (\sin^2
\th - \sin^2 \th_{min} )}}{\sin \th} , \ee where \bea \la{inf}
\sin^2 \th_{max} + \sin^2 \th_{min} &=& \frac{\k^2 (a-b)^2 + 2b\n
c}{b^2 \n^2} ,\nn \sin^2 \th_{max} \cdot \sin^2 \th_{min} &=&
\frac{c^2}{b^2 \n^2} . \eea The second Virasoro constraint is reduced
to a relation among various parameters. From this we can set \be
\la{con1} a = \frac{\n}{\k^2} c . \ee Note that in the above, the
equation of motions are independent of NS-NS $B$ field.
As
previously mentioned, in this system there exists two conserved
charges namely the total energy $E$ and total angular momentum $J$ .
The energy is
given by \bea E &\equiv& \frac{\sqrt{2\l}}{2}
\int d \s \pa_{\ta} t \nn &=&  \sqrt{2\l}
\int_{\th_{min}}^{\th_{max}} d \th  \ \frac{\k (a^2 -b^2)}{b^2 \n}
\frac{\sin \th}{\sqrt{ (\sin^2 \th_{max} - \sin^2 \th) (\sin^2 \th
- \sin^2 \th_{min} )}} \eea and angular momenta is given by \bea J
&\equiv& - \frac{\sqrt{2\l}}{2} \int d \s \sin^2 \th \pa_{\ta} \ph
+ \frac{B}{4\pi}
\int d \s \sin\th \pa_{\s} \th \nn
      &=& - \sqrt{2\l}\int_{\th_{min}}^{\th_{max}} d \th \frac{1}{b^2 \n}
      \frac{\sin \th (ac - b^2 \n \sin^2 \th)}{\sqrt{ (\sin^2 \th_{max} - \sin^2 \th)
(\sin^2 \th - \sin^2 \th_{min} )}
} + \frac{B}{2\pi}
\int_{\th_{min}}^{\th_{max}} d \th \sin\th . \nn  \eea To consider
a giant magnon or spike solution, we have to define the world
sheet momentum $p$, which is identified with the angle difference
$\D \ph \equiv p$,
\bea \la{ang} \D \ph &\equiv& - \int d \ph = - 2
\int_{\th_{min}}^{\th_{max}} d \th \frac{h'}{\th'} \nn &=& - 2
\int_{\th_{min}}^{\th_{max}} d \th \frac{1 }{b^2 \n}
      \frac{(b c - a b \n \sin^2 \th)}{\sin \th \sqrt{ (\sin^2 \th_{max} - \sin^2 \th)
(\sin^2 \th - \sin^2 \th_{min} )}} , \eea where we use a minus
sign for making the angle difference positive.
\section{Giant Magnon and Single Spike Solutions}
In this section we will find the dispersion relation among various
charges defined in the previous section, in the infinite size
limit, which implies infinite angular momentum in case of giant
magnon and infinite angle difference in case of a spiky string
solution. This infinite size limit can be obtained by setting
$\sin \th_{max} = 1$ in both cases. In other words, this limit
reduces \eq{the} to \be \th' = \frac{b \n}{a^2 -b^2} \frac{\cos
\th \sqrt{ (\sin^2 \th - \sin^2 \th_{min} )}}{\sin \th} . \ee One
sees that all of $E, J$ and $\D \phi$ diverge except for a special
parameter region. In this limit, the second line in \eq{inf} gives
$\sin^2 \th_{min} = \frac{c^2}{b^2 \n^2}$. The first line in
\eq{inf} gives rise to a relation among various parameters. Using
this relation and taking the help of \eq{con1}, we get \be
\la{con2} {\k}^2(a^2 -b^2) = (c-b{\n})^2. \ee
{\bf Giant Magnon}\\

To find a solution which correspond to giant magnon, we first have
to use a special parameter region which will make both $E-J$ and
$\D\phi$ finite.

Below we summarize some details of identifying the parameter region. First
we see that to cancel the logarithmic divergence in
$ E-J $ we get the following relation
\bea
\k (a^2-b^2) = (ac -b^2 \n). \eea
If we further demand the finiteness of $\Delta \phi $, then we get
\bea
c + 2ab\n = 0 .
\eea
Using the relations among other parameters as mentioned in the
previous section, we find that
\bea
\n = -\k ,~~ a = -\frac{c}{\k}.
\eea
which makes
\bea \sin^2 \theta_{min}= \frac{a^2}{b^2}. \eea With the above,
one finds that making $ E -J $ finite,  $\Delta \phi$
automatically becomes finite and vice versa. The expression for
$E-J$ and $\Delta \phi$ now becomes \bea \la{rEJ} E - J &=&
\left(\sqrt{2\l} - \frac{B}{2\pi}\right) z_{max} , \nn \D\ph &=&
2\left(\frac{\pi}{2}-\th_{min}\right) , \eea where $z_{max} = \cos
\theta_{min}$. Finally we get the following dispersion relation
for the giant magnon solution in the presence of NS-NS  $B$ field
as \bea E-J = \left(\sqrt{2\l} - \frac{B}{2\pi}\right)\left|\sin
\frac{\D\phi}{2}\right| . \eea
Note that for $B=0$, we get back the dispersion relation obtained in \cite{GHO,GGY}.\\
{\bf Single Spike}\\
In order to find a spike solution, we impose that $J$ is finite.
We also impose that $E  - \frac{\sqrt{2 \l}}{2} \Delta \phi $ is
finite. So one finds following constraints on various parameters:
\bea ac = b^2 \n, ~~ a = \frac{\n}{\k^2}c , \eea including those
we have from the Virasoro constraints previously.
A consistent parameter region can be summarized as follows \bea a
= -\frac{\n}{\k}, ~~ c = -\k, ~~~~ \sin \theta_{min} =
\frac{\k}{\n}. \eea

Finally one can compte the dispersion relation for the spike
solution as \bea E - {\sqrt{2\l}\over 2} \Delta \phi = \sqrt{2\l}
~~\bar{\theta}. \nn J = \left(\sqrt{2\l} - \frac{B}{2\pi}\right)
\sin \bar{\theta} \eea where $\bar{\theta} = \pi/2 -
\theta_{min}$. We further note that for $B=0$, the above relation
reduces to that of \cite{LPP}.

\section{Finite Size effects}
We investigated giant magnon and spike solutions for the
string in the infinite size limit in the last section.In this section we investigate the finite
size effect on them.
For that purpose we take $\th_{max} \ne \pi/2$. \vskip .1in \noindent
{\bf Giant magnon:} \\
\eq{inf} and \eq{con1} gives
\bea \sin \th_{max} &=& - \frac{\k}{\n} , \nn \sin
\th_{min} &=& \ \frac{c}{\k b} , \eea 
Assuming $z \equiv \cos \th$,conserved
charges are given by \bea    \la{deeq} E &=& \sqrt{2\l} \
\frac{z_{max}^2 - z_{min}^2}{z_{max} \sqrt{1-z_{min}^2}} \ K(x),
\nn J &=& \sqrt{2\l} z_{max} \  \lb  K(x) - E(x) \rb  +\frac{B}{2\pi}(z_{max}-z_{min}) , \nn \frac{\D
\ph}{2} &=& \frac{\sqrt{1-z_{min}^2}}{z_{max} \sqrt{1-z_{max}^2}}
\Pi \ls \frac{z_{max}^2 - z_{min}^2}{\sqrt{z_{max}^2-1}};x \rs
 - \frac{\sqrt{1-z_{max}^2}}{z_{max} \sqrt{1-z_{min}^2}}  K(x),
\eea where we have used $z_{max}^2 \equiv \cos^2 \th_{min}=
\frac{k^2 b^2 -c^2}{\k^2 b^2}$, $z_{min}^2 \equiv \cos^2 \th_{max}
= \frac{\n^2 - \k^2}{\n^2}$,$x = \sqrt{1-
\frac{z_{min}^2}{z_{max}^2}}$ and the elliptic integrals of the first, second and
third kinds \bea K(x) &=&  \int_{z_{min}}^{z_{max}} dz
\frac{z_{max}}{\sqrt{(z_{max}^2 - z^2)(z^2 - z_{min}^2)}}  , \nn
E(x) &=&  \int_{z_{min}}^{z_{max}} dz \frac{z^2}{z_{max}
\sqrt{(z_{max}^2 - z^2)(z^2 - z_{min}^2)}} , \nn \Pi \ls
\frac{z_{max}^2 - z_{min}^2}{\sqrt{z_{max}^2-1}};x \rs &=&
 \int_{z_{min}}^{z_{max}} dz
\frac{ z_{max} (1-z_{max}^2)}{(1-z^2) \sqrt{(z_{max}^2 - z^2)(z^2
- z_{min}^2) }} . \eea 
To find the finite size effect,
expanding conserved charges to ${\cal O} (z_{min}^2)$ and
${\cal O} (z_{max}^2)$, we obtain
\bea
\la{fe1} E-J &\approx&
\sqrt{2\l} \ls \left| \sin \frac{\D\phi}{2} \right| - \frac{z_{max}
z_{min}^2}{4}\rs +\frac{B}{2\pi}(z_{min} - z_{max}). \eea
The leading behaviors of $E$ and $z_{max}$ are given by
\bea
\la{ere}
E &\approx & \sqrt{2\l} z_{max} \log \frac{4 z_{max}}{z_{min}} ,
\nn
z_{max} &\approx &  \left|\sin \frac{\D\phi}{2} \right|,
\nn
z_{min} &\approx &   \left| 4\sin \frac{\D\phi}{2} \right|
e^{- E /(\sqrt{2 \l}|\sin \frac{\D\phi}{2})}. \eea
Using above relations,one obtains the following
dispersion relation for giant magnon solution including the
finite size correction as \be E - J
=  \sqrt{2 \l} \left(\left|  \sin \frac{\D\phi}{2} \right| - 4 \left|
\sin^3 \frac{\D\phi}{2} \right| e^{- E / ( \sqrt{2 \l} \left|  \sin
\frac{\D\phi}{2} \right| )}\right)+\frac{B}{2\pi}\left| 4\sin \frac{\D\phi}{2} \right|( e^{- E / ( \sqrt{2 \l} \left| \sin
\frac{\D\phi}{2} \right|)}  - \frac{1}{4})  . \ee For the infinite size case both
$E \ {\rm and} \ J \to \infty$, which gives exactly same result
obtained in the previous section.\vskip .1in
\noindent
{\bf Single Spike:}\\
In this section we calculate finite size effect for the spike solution
obtained in the previous section.
Note that for the angular momentum to be positive in case of the
spike solution, we should consider $J' \equiv - J$ by changing the
directions of rotation. Keeping this in mind, we now start to
calculate the finite size effect for the spike.

Before calculating the dispersion relation, we should find $\sin \th_{min}$
and $\sin \th_{max}$, which are given by
\bea
\sin \th_{min} &\equiv&  \sqrt{1-z_{max}^2} =  \frac{\k}{\n}  , \nn
\sin \th_{max} &\equiv&  \sqrt{1-z_{min}^2} =  \frac{c}{\k b}  .
\eea
With the above, the conserved charges can be rewritten as
\bea
E &=& \sqrt{2\l}  \ \frac{  z_{max}^2 - z_{min}^2  }{z_{max} \sqrt{1-z_{max}^2}} \ K(x)  ,\nn
J' &=& \sqrt{2\l}\ \frac{1}{z_{max}} \ls  z_{max}^2 E (x) - z_{min}^2 K(x) \rs - \frac{B}{2\pi}(z_{max} - z_{min}), \nn
\frac{\D \ph}{2} &=& \frac{\sqrt{1-z_{min}^2}}{z_{max} \sqrt{1-z_{max}^2}} \lb K(x)
- \Pi \ls \frac{z_{max}^2 - z_{min}^2}{\sqrt{z_{max}^2-1}};x \rs \rb .
\eea

Here, the angular
momentum $J'$ is given by
\be   \la{jeq}
J' \approx \sqrt{2\l} z_{max}  - \ls \half + \log \frac{4 z_{max}}{z_{min}} \rs
\frac{\sqrt{2\l} z_{min}^2}{2z_{max}} - \frac{B}{2\pi}(z_{max} - z_{min}) ,
\ee
at ${\cal O} (z_{min}^2)$. Note that for the infinite size limit, $z_{min} \to 0$,
the second term in the right hand side vanishes, so $J'$ is always finite as it should do.
The dispersion relation for a spike $E- \frac{\sqrt{2\lambda}}{2}\D \ph$
is given by
up to ${\cal O}(z_{min}^3)$ and ${\cal O}(z_{max}^3)$
\bea    \la{predis}
E-\frac{\sqrt{2\l}}{2}\D \ph &\approx & \sqrt{2\l} \arcsin z_{max} \nn
  & - & \lb  \ls \frac{1}{2 z_{max}} - \frac{ z_{max}}{4} \rs \frac{\sqrt{2\l}}{2}
 + \ls \frac{1}{z_{max}} + \frac{z_{max}}{2} \rs \frac{\sqrt{2\l}}{2} \log \frac{4 z_{max}}{z_{min}} \rb z_{min}^2 .\nn
\eea
To rewrite the dispersion relation in terms of the physical quantities,
$z_{min}$ and $z_{max}$ should be replaced with $E$ and $J'$. From the leading term
of $E$
we obtain
\be
z_{min} = 4 z_{max} e^{-E/\sqrt{2\l} z_{max}}
\ee
and the leading term of $J'$ gives
\be
z_{max} = \frac{J'}{(\sqrt{2\l} - \frac{B}{2\pi})}\equiv {\cal J}(B) .
\ee

So
\be
z_{min} = 4 z_{max} e^{-E/\sqrt{2\l} z_{max}} = 4{\cal J}(B) e^{-E/\sqrt{2\l}{\cal J}(B) }
\ee

Using these, we finally obtain the dispersion relation for a finite size spike solution
\bea   \la{edis}
E - \frac{\sqrt{2\l}}{2} \D \ph & \approx  &\sqrt{2\l} \arcsin {\cal J}(B)    \nn
&-& 8 \left(\sqrt{1- ({\cal J}(B))^2} \frac {\sqrt{2\l} {\cal J}(B)} {2}
+ \frac{E}{\sqrt{1- ({\cal J}(B)})^2}\right)
    e^{-2 E /\sqrt{2\l} {\cal J}(B)  } .\nn
\eea

\section{Conclusions} We have studied, in this paper, the rotating
string in the diagonal $SU(2)$ subspace inside the AdS$_4 \times
{\bf CP}^3$ geometry in the presence of a NS-NS $B$ field. We see
that although the equations of motion of the rotating string on
$R\times S^2$ are independent of NS-NS $B$ field, the most general
form of conserved angular momentum depends on $B$ field. We have
shown the existence of both the giant magnon, and the spike
solutions for the string moving in this background which
correspond to open spin chain and have found out the relevant
dispersion relation among various charges in the infinite size
limit. Furthermore, we have studied the finite size correction in
both cases. It will be interesting to find more general string
solutions in this background following \cite{RNG}.


\newcommand{\np}[3]{Nucl. Phys. {\bf B#1}, #2 (#3)}
\newcommand{\pprd}[3]{Phys. Rev. {\bf D#1}, #2 (#3)}
\newcommand{\jjhep}[3]{J. High Energy Phys. {\bf #1}, #2 (#3)}


\begin{thebibliography}{20}
\bibitem{ABJM} O. Aharony, O. Bergman, D.L. Jafferis and J. Maldacena,
``$\mathcal{N}=6$ superconformal Chern-Simons-matter theories,
M2-branes and their gravity duals," arXiv:0806.1218[hep-th].
JHEP \textbf{08} (2008) 001 [arXiv:0806.3391[hep-th]].
\bibitem{Aharony:2008gk}
O.~Aharony, O.~Bergman and D.~L.~Jafferis,
``Fractional M2-branes,''
arXiv:0807.4924 [hep-th].
\bibitem{MGW} J.M. Maldacena, ``The large N limit of superconformal
field theories and supergravity," Adv. Theor. Math. Phys. \textbf{2}
(1998) 231 [arXiv:hep-th/9711200]; S.S. Gubser, I.R. Klebanov and
A.M. Polyakov, ``Gauge theory correlators from non-critical string
theory," Phys. Lett. \textbf{B428} (1998) 105 [arXiv:hep-th/9802109];
 E. Witten, ``Anti-de Sitter space and holography," Adv. Theor.
Math. Phys. \textbf{2} (1998) 253 [arXiv:hep-th/9802150].

\bibitem{HM} D.M. Hofman and J.M. Maldacena, ``Giant magnons,"
J. Phys. \textbf{A39} (2006) 13095 [arXiv:hep-th/0604135].
\bibitem{Kruczenski:2004wg}
M.~Kruczenski,
JHEP {\bf 0508}, 014 (2005)
[arXiv:hep-th/0410226].
\bibitem{krt0607}
 M. Kruczenski, J. Russo and A.A. Tseytlin,
\jjhep{0610}{002}{2006} [arXiv:hep-th/0607044].
\textbf{D76} (2007) 126006 [arXiv:0705.2429[hep-th]].

\bibitem{BIKS}
M. Benna, I. Klebanov, T. Klose and M. Smedback, ``Superconformal
Chern-Simons theories and $AdS_4/CFT_3$ correspondence,"
JHEP \textbf{09} (2008) 072 [arXiv:0806.1519[hep-th]].

\bibitem{MZ} J.A. Minahan and K. Zarembo, ``The Bethe ansatz for
superconformal Chern-Simons," JHEP \textbf{09} (2008) 040
[arXiv:0806.3951[hep-th]].

\bibitem{GGY} D. Gaiotto, S. Giombi and X. Yin, ``Spin chains in
$\mathcal{N}=6$ superconformal Chern-Simons-matter theory,"
arXiv:0806.4589[hep-th].

\bibitem{GHO} G. Grignani, T. Harmark and M. Orselli, ``The
$SU(2) \times SU(2)$ sector in the string dual of $\mathcal{N}=6$
superconformal Chern-Simons theory," arXiv:0806.4959[hep-th].

\bibitem{Bak:2008cp}
  D.~Bak and S.~J.~Rey,
  ``Integrable Spin Chain in Superconformal Chern-Simons Theory,''
  JHEP {\bf 0810}, 053 (2008)
  [arXiv:0807.2063 [hep-th]].

\bibitem{AF} G. Arutyunov and S. Frolov, ``Superstrings on
$AdS_4 \times CP^3$ as a coset sigma-model," arXiv:0806.4940[hep-th];
B.J. Stefanski, ``Green-Schwarz action for type IIA strings on
$AdS_4 \times CP^3$," arXiv:0806.4948[hep-th];
  G.~Bonelli, P.~A.~Grassi and H.~Safaai,
``Exploring Pure Spinor String Theory on AdS(4) X CP**3,''
 arXiv:0808.1051 [hep-th].

\bibitem{Gromov:2008qe}
   N.~Gromov and P.~Vieira,
   ``The all loop AdS4/CFT3 Bethe ansatz,''
   arXiv:0807.0777 [hep-th].

\bibitem{NGV} N. Gromov and P. Vieira, ``The $AdS_4/CFT_3$
algebraic curve," arXiv:0807.0437[hep-th].
\bibitem{LPP} B.H. Lee, K.L. Panigrahi and C. Park, ``Spiky strings
on $AdS_4 \times CP^3$," arXiv:0807.2559[hep-th].
\bibitem{CW} B. Chen and J.B. Wu, ``Semi-classical strings
in $AdS_4 \times CP^3$," arXiv:0807.0802[hep-th].
\bibitem{RNG} Shijong Ryang,
"Giant Magnon and Spike Solutions with Two Spins in AdS4xCP3,"
arXiv:0809.5106 [hep-th].
\bibitem{ABR} C. Ahn, P. Bozhilov and R.C. Rashkov,
``Neumann-Rosochatius integrable system for strings
on $AdS_4 \times CP^3$," JHEP \textbf{09} (2008) 017
[arXiv:0807.3134[hep-th]].
\bibitem{IS} I. Shenderovich, ``Giant magnons in $AdS_4/CFT_3$:
dispersion, quantization and finite-size corrections,"
arXiv:0807.2861[hep-th].
\bibitem{MR} T. McLoughlin and R. Roiban, ``Spinning strings
at one-loop in $AdS_4 \times P^3$," arXiv:0807.3965[hep-th].
\bibitem{AAB} L.F. Alday, G. Arutyunov and D. Bykov, ``Semiclassical
quantization of spinning strings in $AdS_4 \times CP^3$,"
arXiv:0807.4400[hep-th].
\bibitem{CK} C. Krishnan, ``$AdS_4/CFT_3$ at one loop,"
arXiv:0807.4561[hep-th].
\bibitem{GM} N. Gromov and V. Mikhaylov, ``Comment of the scaling
function in $AdS_4 \times CP^3$," arXiv:0807.4897[hep-th].
\bibitem{BGR} D. Bak, D. Gang and S.J. Rey, ``Integrable spin
chain of superconformal $U(M)\times \bar{U}(N)$ Chern-Simons
theory," arXiv:0808.0170[hep-th].

\bibitem{IK} R. Ishizeki and M. Kruczenski, ``Single spike solutions
for strings on $S^2$ and $S^3$," Phys. Rev.
\bibitem{NBR} N.P. Bobev and R.C. Rashkov, ``Spiky strings, giant
magnons and beta-deformations," Phys. Rev. \textbf{D76} (2007)
046008 [arXiv:0706.0442[hep-th]];
Hayashi, K. Okamura, R. Suzuki and B. Vicedo,
``Large winding sector of AdS/CFT," JHEP \textbf{11} (2007) 033
[arXiv:0709.4033[hep-th]];
H. Dimov and R.C. Rashkov, ``On the anatomy of multi-spin magnon
and single spike string solutions," Nucl. Phys. \textbf{B799}
(2008) 255 [arXiv:0709.4231[hep-th]].
\bibitem{KNP} J. Kluson, R.R. Nayak and K.L. Panigrahi, ``Giant
magnon in NS5-brane background," JHEP \textbf{04} (2007) 099
[arXiv:hep-th/0703244];B.H. Lee, R.R. Nayak, K.L. Panigrahi and C. Park, ``On the giant
magnon and spike solutions for strings on $AdS_3 \times S^3$,"
JHEP \textbf{06} (2008) 065 [arXiv:0804.2923[hep-th]];
J.R. David and B. Sahoo, ``Giant magnons in the D1-D5 system,"
JHEP \textbf{07} (2008) 033 [arXiv:0804.3267[hep-th]];
J. Kluson, B.H. Lee, K.L. Panigrahi and C. Park, ``Magnon like solutions
for strings in I-brane background," JHEP \textbf{08} (2008)
032 [arXiv:0806.3879[hep-th]];
C.M. Chen, J.H. Tsai and W.Y. Wen, ``Giant magnons and spiky strings
on $S^3$ with B-field," arXiv:0809.3269[hep-th];J. Kluson,K.L. Panigrahi,
"D1-brane in beta-Deformed Background,"JHEP \textbf{0711}:011,(2007) arXiv:0710.0148 [hep-th].

\bibitem{AFZ} G. Arutyunov, S. Frolov and M. Zamaklar, ``Finite-size
effects from giant magnons," Nucl. Phys. \textbf{B778} (2007) 1
[arXiv:hep-th/0606126];D. Astolfi, V. Forini, G. Grignani and G.W.
Semenoff, ``Gauge invariant finite size spectrum of the
giant magnon," Phys. Lett. \textbf{B651} (2007) 329
[arXiv:hep-th/0702043];
Y. Hatsuda and R. Suzuki, ``Finite-size effects for
dyonic giant magnons," Nucl. Phys. \textbf{B800} (2008) 349
[arXiv:0801.0747[hep-th]];
``Finite-size effects for multi-magnon states,"
arXiv:0807.0643[hep-th];C. Ahn and P. Bozhilov, ``Finite-size effects for single spike,"
JHEP \textbf{07} (2008) 105 [arXiv:0806.1085[hep-th]];
T. Klose and T. McLoughlin, ``Interacting finite-size magnons,"
J. Phys. \textbf{A41} (2008) 285401 [arXiv:0803.2324[hep-th]];
B. Ramadanovic and G.W.Semenoff, ``Finite size giant magnon,"
arXiv:0803.4028[hep-th];G. Grignani, T. Harmark, M. Orselli and G.W. Semenoff,
``Finite size giant magnons in the string dual of $\mathcal{N}=6$
superconformal Chern-Simons theory," arXiv:0807.0205[hep-th];
D. Astolfi, V.G.M. Puletti, G. Grignani, T. Harmark and M. Orselli,
``Finite-size corrections in the $SU(2) \times SU(2)$ sector of
type IIA string theory on $AdS_4 \times CP^3$,"
arXiv:0807.1527[hep-th].
Changrim Ahn, P. Bozhilov,
" Finite-size Effect of the Dyonic Giant Magnons in N=6 super Chern-Simons Theory,"
arXiv:0810.2079 [hep-th];Tomasz Lukowski, Olof Ohlsson Sax,
"Finite size giant magnons in the $SU(2) x SU(2)$ sector of $AdS_4 x CP^3,$ "
arXiv:0810.1246 [hep-th];Diego Bombardelli, Davide Fioravanti,
" Finite-Size Corrections of the ${CP}^3$ Giant Magnons: the Luscher terms,"
arXiv:0810.0704 [hep-th]; J. A. Minahan and O. Ohlsson Sax,
"Finite size effects for giant magnons on physical strings",
Nucl. Phys.  B801 (2008) 97
[arXiv:0801.2064 [hep-th]];
   N.~Gromov, S.~Schafer-Nameki and P.~Vieira,
   ``Quantum Wrapped Giant Magnon,''
   Phys.\ Rev.\  D {\bf 78} (2008) 026006
   [arXiv:0801.3671 [hep-th]];
   N.~Gromov, S.~Schafer-Nameki and P.~Vieira,
   ``Efficient precision quantization in AdS/CFT,''
   arXiv:0807.4752 [hep-th].



\end{thebibliography}
\end{document}